\documentclass[
    ,final            
  ]
  {aipproc}

\layoutstyle{8x11single}

\begin{document}

\title{Electroweak Physics: Theoretical Overview~\footnote{
Talk presented at the {\em Hadron Collider Physics Symposium 2006} at Duke University, Durham, North Carolina.}}


\author{Doreen Wackeroth}{
address={Center for Particles and Fields and Department of Physics, University of Texas, Austin, TX 78712, USA\footnote{On leave from the Department of Physics, SUNY at Buffalo, Buffalo, NY 14260, USA. Electronic address: dow@ubpheno.physics.buffalo.edu.}}}


\begin{abstract}
I give an overview of the theory status of predictions for single
$W$ and $Z$ boson production at hadron colliders.  I briefly
report on work in progress for improvements necessary to match the
anticipated high precision of electroweak measurements, such as the
$W$ mass and width, at the Fermilab Tevatron $p\bar p$ and the CERN
LHC $pp$ colliders.
\end{abstract}

\maketitle

\section{Introduction}
\begin{figure}\label{fig:one}
\includegraphics[height=.25\textheight]{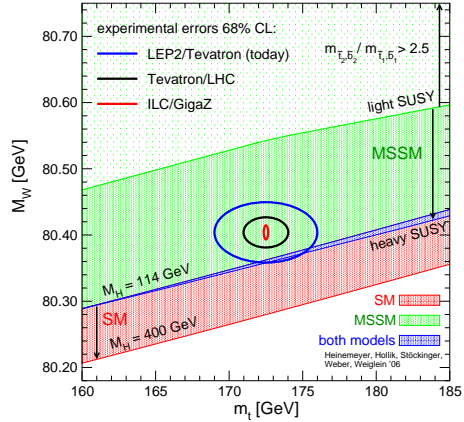}
\caption{The predictions for the $W$ mass in dependence of the top
quark mass within the SM and MSSM in comparison with experimental
$M_W$ and $m_t$ measurements at 68\% C.L.. The bands are obtained by
varying the free parameters of the underlying model. Taken from Ref.~\cite{Heinemeyer:2006px}. }
\end{figure}

\begin{figure}\label{fig:two}
\includegraphics[height=.2\textheight]{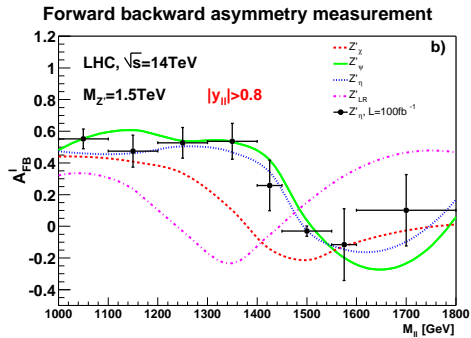}
\caption{The forward-backward asymmetry, $A_{FB}(M(ll))$, 
of single $Z'$ production in $pp\to Z'\to l^+l^-$ at the LHC for 
a number of models with heavy, non-standard gauge bosons. Taken from Ref.~\cite{Dittmar:2003ir}.}
\end{figure}

Electroweak gauge boson production processes are one of the best, most
precise probes of the Standard Model (SM). The electroweak physics
program involving single $W$ and $Z$ boson production at hadron
colliders has many facets:
\begin{itemize}
\item
The comparison of direct measurements of the $W$ boson mass ($M_W$)
and width ($\Gamma_W$) in $W$ pair production at LEP2 and single $W$
production at the Tevatron, with indirect measurements from a global
fit to electroweak precision data measured at LEP1/SLD, represents a
powerful test of the SM.  Any disagreement could be interpreted as
a signal of physics beyond the SM.  At present, direct and indirect
measurements of $M_W$ and $\Gamma_W$ agree within their respective
errors~\cite{lepewwg}: $M_W$(LEP2/Tevatron)$=80.392\pm 0.029$ GeV
versus $M_W$(LEP1/SLD)$=80.363\pm 0.032$ GeV and
$\Gamma_W$(LEP2/Tevatron)$=2.147\pm 0.060$ GeV versus
$\Gamma_W$(LEP1/SLD)$=2.091\pm 0.003$ GeV. Continued improvements in
theory and experiment will further scrutinize the SM.
\item
The precise measurements of $M_W$ and the top quark mass ($m_t$)
provide an indirect measurement of the SM Higgs boson mass, $M_H$, and
a window to physics beyond the SM, as illustrated in
Figure~\ref{fig:one}~\cite{Heinemeyer:2006px}. With the present
knowledge of $M_W$ and $m_t$~\cite{lepewwg,mtop}, the SM Higgs boson
mass can be indirectly constrained by a global fit to all electroweak
precision data to be smaller than 199 GeV at 95 \%
C.L.~\cite{lepewwg}.  Future more precise measurements of $M_W$ and
$m_t$ will considerably improve the present indirect bound on $M_H$:
At the LHC, for instance, with anticipated experimental precisions of
$\delta M_W=15$~MeV and $\delta m_t=1$~GeV, $M_H$ can be predicted
with an uncertainty of about $\delta
M_H/M_H=18\%$~\cite{Baur:2002gp}. In
Figure~\ref{fig:one}~\cite{Heinemeyer:2006px} the predictions for
$M_W(m_t,M_H,\ldots)$ within the SM and the minimal supersymmetric SM
(MSSM) are confronted with their measurements today, at the LHC, and
an International Linear Collider (ILC).
\item
The measurement of the mass and width of the $Z$ boson and the total
$W$ and $Z$ production cross sections can be used for detector
calibration and as luminosity monitors~\cite{Dittmar:1997md},
respectively.
\item
The $W$ charge asymmetry and $Z$ rapidity distributions severely
constrain quark Parton Distribution Functions (PDFs).
\item
New, heavy gauge bosons may leave their footprints in forward-backward
asymmetries, $A_{FB}$, and the distribution of the invariant mass of the lepton
pair, $M(ll)$, produced in $Z$ boson production at high $M(ll)$. In
Figure~\ref{fig:two}~\cite{Dittmar:2003ir} the effects of a $Z'$ on
$A_{FB}(M(ll))$ at the LHC are shown, assuming a number of different
models of extended gauge boson sectors, and compared with simulated
data assuming a specific model. As can be seen, measurements of
$A_{FB}$ at the LHC will be able to distinguish between different new
physics scenarios provided, of course, the SM prediction is well under
control.
\end{itemize}

\begin{figure}
  \unitlength 1cm
  \begin{picture}(12.,4.5)
  \put(-3.,-6.8){\includegraphics[width=9.5cm]{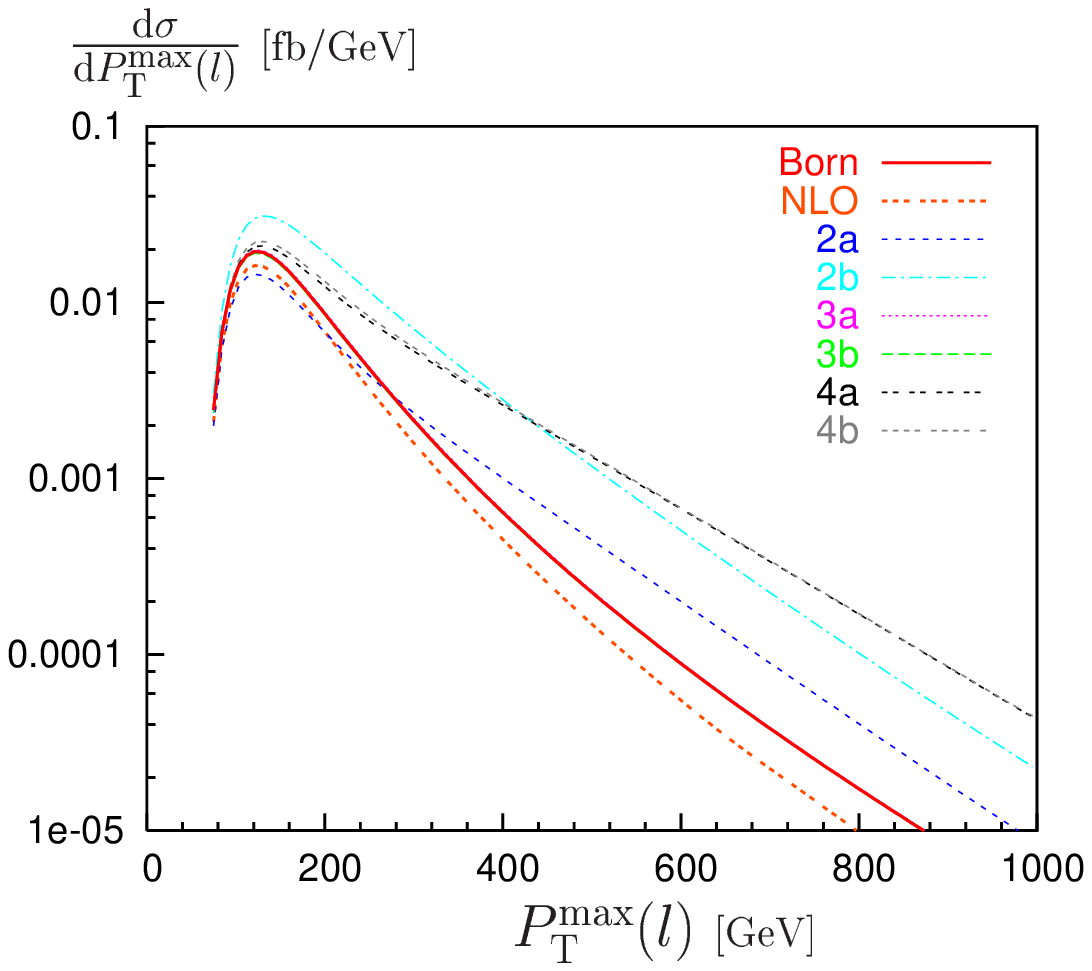}}
  \put(5.,-6.8){\includegraphics[width=9.5cm]{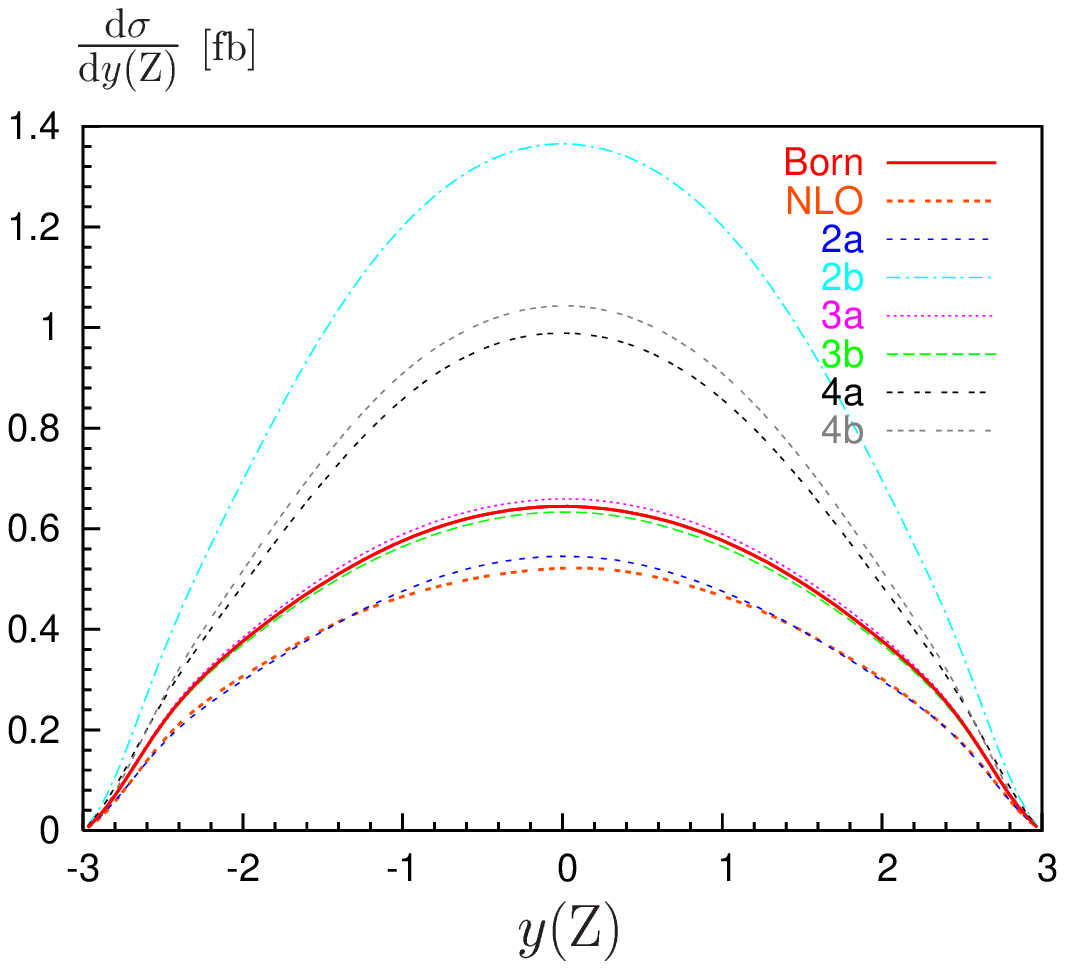}}
  \end{picture}
\caption{The maximal transverse momentum distribution of the charged
leptons and the rapidity distribution of the reconstructed $Z$ boson
in $WZ$ production at the LHC. Shown is the lowest order (Born) and
the next-to-leading order (NLO) result including electroweak
corrections. The labels $2a-4b$ denote different choices for anomalous
triple gauge-boson couplings. For more details see Ref.~\cite{Accomando:2005xp}.}
\label{fig:three}
\end{figure}

In order to fully exploit the potential of the Tevatron and LHC for
electroweak (EW) precision physics, the predictions have to be of the
highest standards as well. The omission of EW radiative corrections in
the comparison of predictions with data could result in fake signals
of non-standard physics. For instance, in Ref.~\cite{Baur:2004ig} it
has been shown that the effects of weak non-resonant corrections on
the tail of the transverse mass distribution of the lepton pair, $M_T(l\nu)$, 
produced in $p\bar p \to W\to l\nu$ at the Tevatron, from which
$\Gamma_W$ can be extracted, are of the same order of magnitude as
effects due to non-SM values of the $W$ width. Another example is $WZ$
production at the LHC, which is a sensitive probe of the non-abelian
structure of the SM EW sector.  As can be seen in
Figure~\ref{fig:three}~\cite{Accomando:2005xp}, effects of non-standard
weak gauge boson self-couplings can be similar in size and shape to
the effects of EW corrections, and, thus, not including the latter
could be mistaken as signals of new physics.
Consequently, in recent years a lot of theoretical effort has gone into
improving the predictions for $W$ and $Z$ production processes in
order to match (or better exceed) the anticipated experimental
accuracy. This not only requires the calculation of higher-order
corrections but also their implementation in Monte Carlo (MC)
integration programs for realistic studies of their effects on
observables. In the following, I will first review some of the
available theoretical tools and then provide a brief overview of the
effects of EW radiative corrections on $W$ and $Z$ boson
observables at hadron colliders. Finally, I will describe recent
developments and work in progress in the context of the TeV4LHC
workshop.

\section{Theoretical Status}
\begin{figure}\label{fig:four}
\includegraphics[height=.2\textheight]{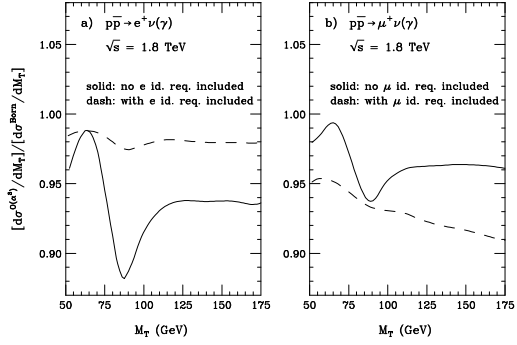}
\caption{The ratio of the full ${\cal O}(\alpha^3)$ and Born
transverse mass distribution of the final-state lepton pair, $M_T(l\nu)$,
in $p\bar p \to W^+ \to l^+ \nu$ ($l=e,\mu$) at the Tevatron with
(dashed line) and without (solid line) lepton identification
requirements.  Shown are the results for the electron and muon case,
which differ significantly in the treatment of the photon emitted
collinear to the charged lepton. For more details see Ref.~\cite{Baur:1998kt}.}
\end{figure}

\begin{figure}\label{fig:five}
\includegraphics[height=.2\textheight]{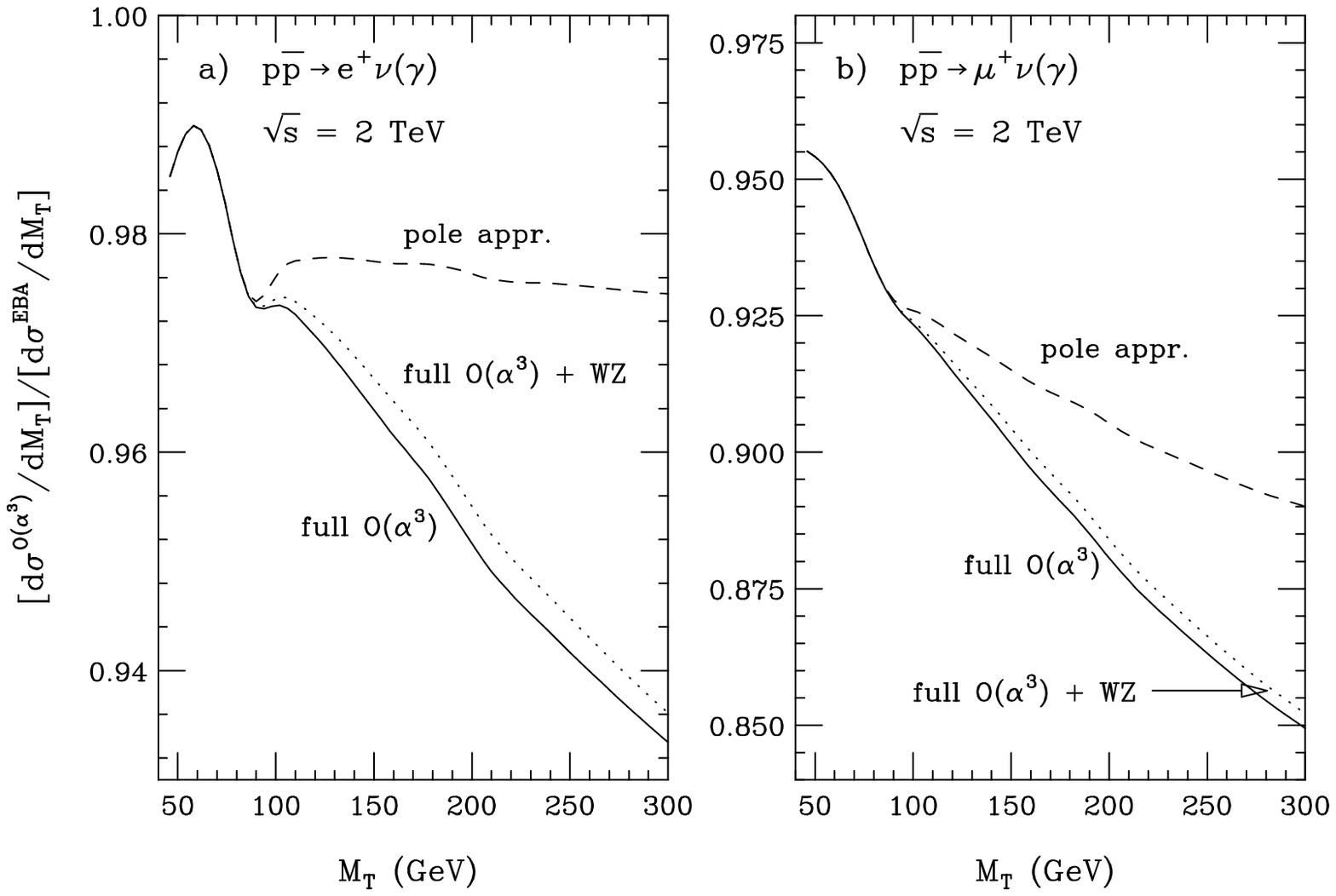}
\includegraphics[height=.2\textheight]{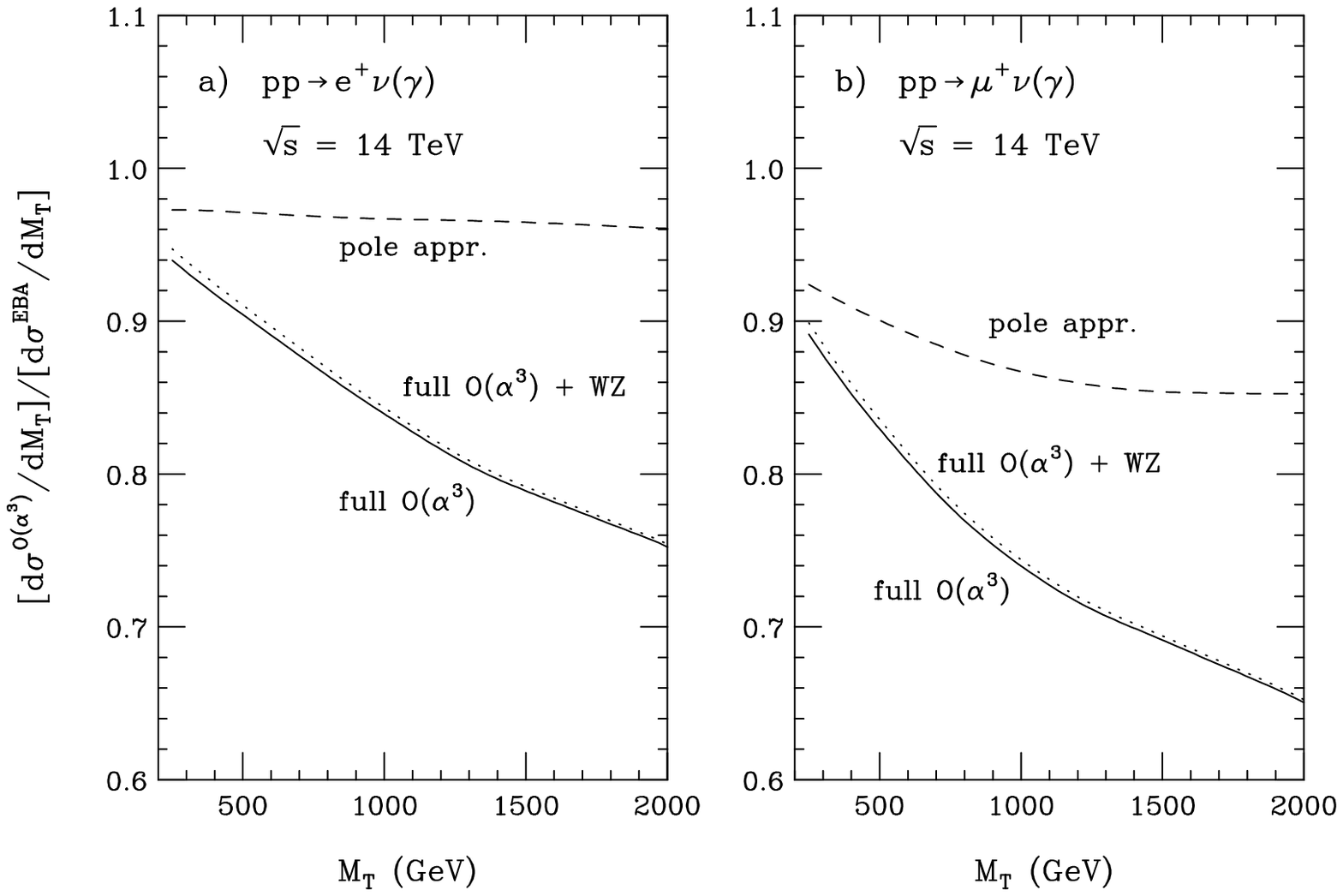}
\caption{The ratio of the ${\cal O}(\alpha^3)$ $M_T(l\nu)$ ($l=e,\mu$)
distribution and an effective Born approximation (EBA) at the Tevatron
and LHC, calculated in the pole approximation (dashed line) and
including the complete EW ${\cal O}(\alpha)$ corrections (solid line).
The difference between these two calculations is mainly due to the
occurrence of large weak Sudakov-like logarithms: they are absent in
the pole approximation where the weak corrections are calculated at
$Q^2=M_W^2$. For more details see Ref.~\cite{Baur:2004ig}.}
\end{figure}

\begin{figure}\label{fig:six}
\includegraphics[height=.2\textheight]{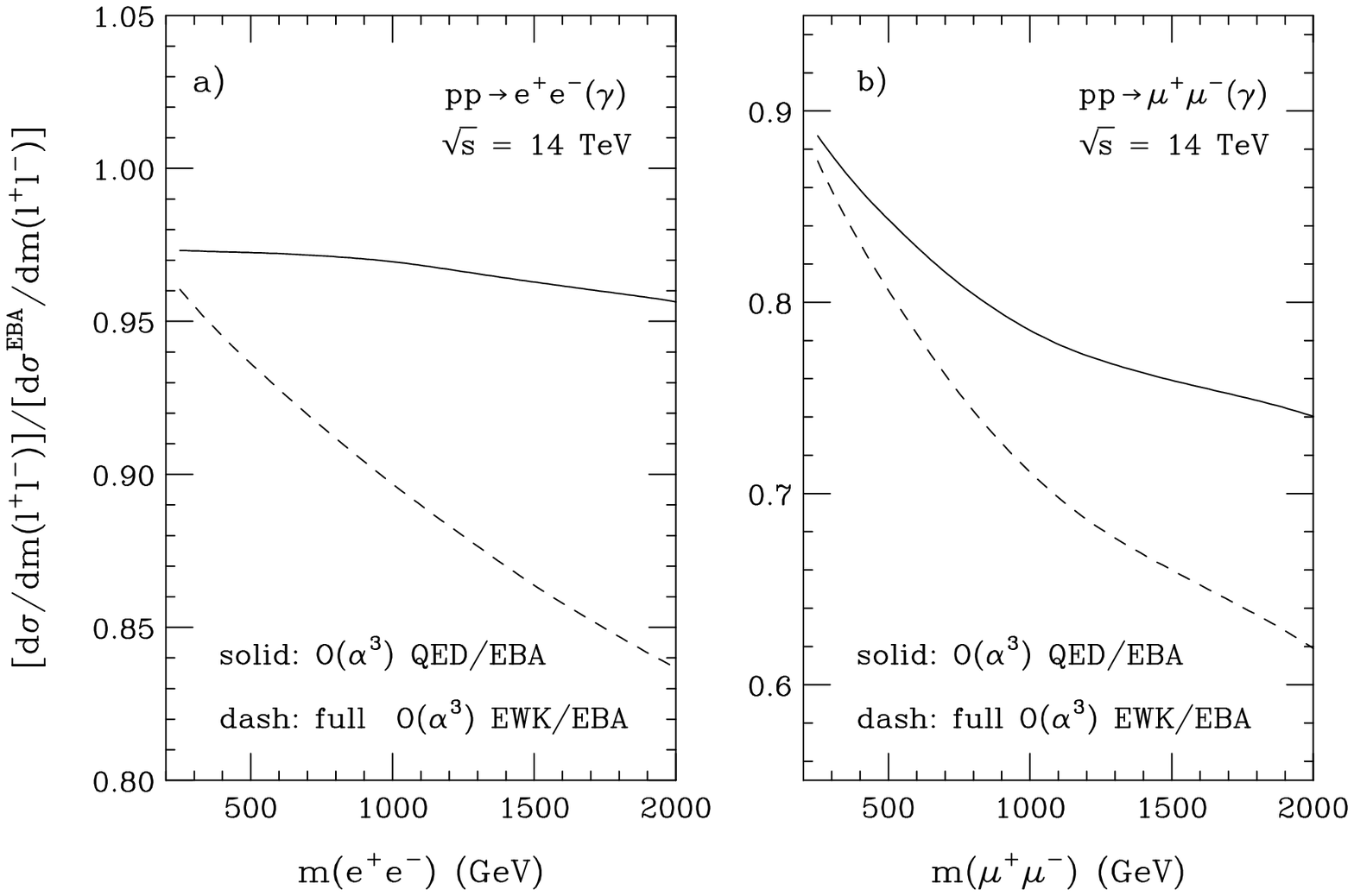}
\includegraphics[height=.2\textheight]{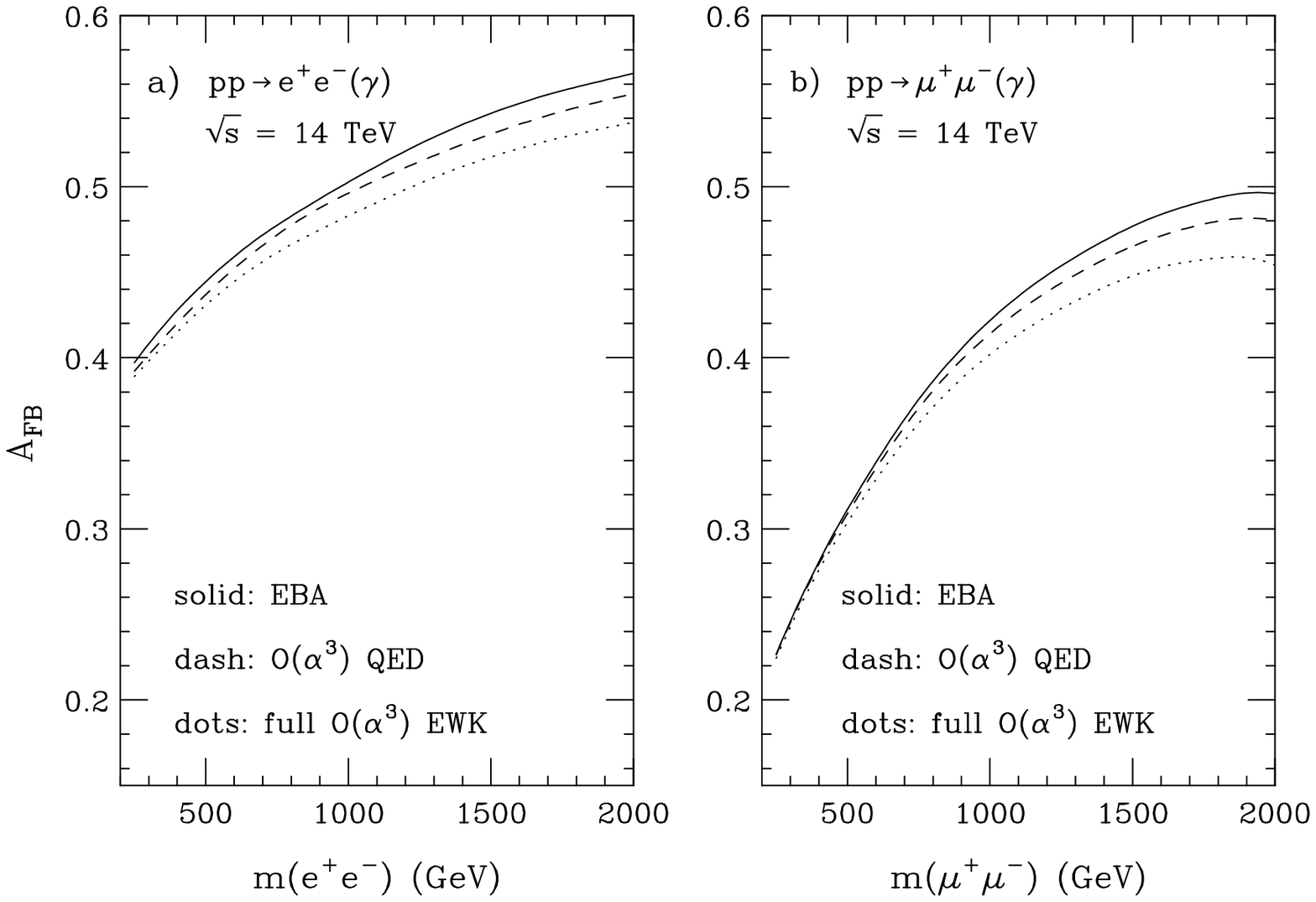}
\caption{Comparison of predictions for the $M_{ll}$ ($l=e,\mu$)
distribution and $A_{FB}(M_{ll})$ that include either only QED or the
complete EW ${\cal O}(\alpha)$ corrections to $Z$ boson production at
the LHC.  The increasing difference between these two calculations
with $M(ll)$ is due to the occurrence of large weak Sudakov-like
logarithms. The effects differ for muons and electrons in the final
state due the differences in the applied lepton identification cuts.
For more details see Ref.~\cite{Baur:2001ze}.}
\end{figure}

Fully differential cross sections for single $W$ and $Z$ boson
production at hadron colliders are known at next-to-next-to-leading
order (NNLO)
QCD~\cite{Anastasiou:2003ds,Anastasiou:2003yy,Melnikov:2006di,Melnikov:2006kv} (and references therein).
Predictions for the $W$ transverse momentum distribution, $p_T(W)$,  
an important
ingredient in the current $W$ mass measurement at the Tevatron,
include an all-order resummation of leading logarithms arising from
soft gluon radiation~\cite{Balazs:1997xd,Ellis:1997ii}.  The complete
EW ${\cal O}(\alpha)$ corrections to $pp,p \bar p \to W
\to l \nu$ and $pp,p \bar p \to Z,\gamma \to l^+ l^-$ have been
calculated in Ref.~\cite{Dittmaier:2001ay,Baur:2004ig,Arbuzov:2005dd,Calame:2006zq}
and~\cite{Baur:2001ze}, respectively. Predictions including multiple
final-state photon radiation have been presented in
Ref.~\cite{CarloniCalame:2003ux,Placzek:2003zg,CarloniCalame:2005vc}.
Most of these higher-order calculations have been implemented in MC
programs and a list of some of the available codes providing precise
prediction for $W$ and $Z$ boson observable at hadron colliders can be
found in Table~\ref{tab:a}.
\begin{table}
\begin{tabular}{ll} \hline
{\tt HORACE}: & Multiple final-state photon radiation in $W$ and $Z$ production as solution of QED DGLAP evolution \\
& for lepton structure functions~\cite{CarloniCalame:2003ux,CarloniCalame:2005vc}, matched with
exact EW ${\cal O}(\alpha)$ corrections to $W$ production~\cite{Calame:2006zq}.\\ 
&{\tt \small http://www.pv.infn.it/$\sim$hepcomplex/horace.html}
\\ \hline
{\tt RESBOS}: & QCD corrections to $W$ and $Z$ production, 
soft gluon resummation, final-state QED ${\cal O}(\alpha)$ corrections~\cite{Balazs:1997xd,Cao:2004yy}.\\
& {\tt \small http://www.pa.msu.edu/$\sim$balazs/ResBos}\\ \hline
{\tt SANC}: & EW corrections to $W$ and $Z$ production: 
automatically generates Fortran code for \\
& one-loop corrections at parton level~\cite{Arbuzov:2005dd,Andonov:2004hi}.
{\tt \small http://sanc.jinr.ru}\\ \hline
{\tt WGRAD2}: & QED ${\cal O}(\alpha)$ and 
weak one-loop corrections to $W$ production~\cite{Baur:2004ig}. \\ 
& {\small \tt http://ubpheno.physics.buffalo.edu/$\sim$dow/wgrad.html} \\
\hline
{\tt WINHAC}:& Multiple final-state photon radiation in
$W$ production 
via YFS exponentiation of soft photons~\cite{Placzek:2003zg}.\\
& {\tt \small http://placzek.home.cern.ch/placzek/winhac}\\
\hline
{\tt ZGRAD2}: & QED ${\cal O}(\alpha)$ and weak one-loop corrections 
to $Z$ production \\
& with proper treatment of higher-order terms around the $Z$ resonance~\cite{Baur:2001ze}.  \\
& {\small \tt http://ubhex.physics.buffalo.edu/$\sim$baur/zgrad2.tar.gz}\\
\hline
\end{tabular}
\caption{MC programs that provide precise predictions
including QED and/or electroweak corrections for $W$ and/or $Z$ boson production at hadron colliders.}
\label{tab:a}
\end{table}
$W$ and $Z$ boson observables are strongly affected by EW corrections.
Their main characteristics can be summarized as follows:
\begin{itemize}
\item 
Photon radiation off the final-state charged lepton can considerably
distort kinematic distributions and usually makes up the bulk of the
effects of EW corrections.  For instance, $W$ and $Z$ boson masses
extracted respectively from the transverse mass and invariant mass
distributions of the final-state lepton pair are shifted by ${\cal
O}(100)$ MeV due to final-state photon radiation.  This is due to the
occurrence of mass singular logarithms of the form $\alpha
\log(Q^2/m_l^2)$ that arise when the photon is emitted collinear to
the charged lepton.  In sufficiently inclusive observables these mass
singularities completely cancel (KLN theorem). But in realistic
experimental environments, depending on the experimental setup, large
logarithms can survive. This is demonstrated in
Figure~\ref{fig:four}~\cite{Baur:1998kt}: the more inclusive
treatment of the photon emitted in $W^+\to e^+ \nu_e$ decays results in a
significant reduction of the final-state QED effects when lepton
identification cuts are applied whereas in the muon case large
logarithms survive.  Because of their numerical importance at
one-loop, the higher-order effects of multiple final-state photon
radiation have to be under good theoretical control as
well~\cite{CarloniCalame:2003ux,Placzek:2003zg,CarloniCalame:2005vc}.
\item
The impact of initial-state photon radiation is negligible after
proper removal of the initial-state mass singularities by universal
collinear counterterms to the quark PDFs.  This mass factorization
introduces a dependence on the QED factorization scheme: in complete
analogy to QCD both the QED DIS and $\overline{\rm MS}$ scheme
have been introduced in the literature~\cite{Baur:1998kt}. Recently,
quark PDFs became available that also incorporate QED radiative
corrections~\cite{Martin:2004dh}, which is important for a consistent
treatment of initial-state photon radiation at hadron colliders.
\item
At high energies, i.e.~in tails of kinematic distributions, for
instance $M(ll) \gg M_Z$ and $M_T(l \nu)\gg M_W$, Sudakov-like
contributions of the form $\alpha \log^2(Q^2/M_V^2)$ ($M_V=M_{W,Z}$
and $Q$ is a typical energy of the scattering process) can
significantly enhance the EW one-loop corrections.  These corrections
originate from remnants of UV singularities after renormalization and
soft and collinear initial-state and final-state radiation of virtual
and real weak gauge bosons.  In contrast to QED and QCD the 
Bloch-Nordsiek theorem is violated, i.e.~even in fully inclusive observables these
large logarithms are present due to an incomplete cancellation between contributions
from real and virtual
weak gauge boson radiation~\cite{Ciafaloni:2000df}. Moreover, the $W$ and $Z$ boson
masses serve as physical cut-off parameters and real $W$ and $Z$ boson
radiation processes are usually not included, since they result in
different initial and/or final states.  The EW logarithmic corrections
of the form $\alpha^L \log^N(\frac{Q^2}{M_V^2}), 1 \le N \le 2L$
($L=1,2 \ldots$ for 1-loop,2-loop,$\ldots$) to 4-fermion processes are
known up to 2-loop $N^3LL$ order and are available in form of compact
analytic formulae (see, e.g.,
Refs.~\cite{Melles:2001ye,Jantzen:2005az,Denner:2006jr} and references
therein). Examples of the effects of these large weak corrections on
$W$ and $Z$ boson observables at one-loop order are shown in
Figure~\ref{fig:five} ($M_T(l\nu)$ distribution)~\cite{Baur:2004ig}
and Figure~\ref{fig:six} ($M(ll)$ and $A_{FB}(M_{ll})$
distributions)~\cite{Baur:2001ze}, respectively.
\end{itemize}
The importance of fully understanding and controlling EW radiative
corrections to precision $W$ and $Z$ boson observables at hadron
colliders is illustrated in Table~\ref{tab:b} on the example of a
precise $W$ mass and width measurement. It demonstrates how
theoretical progress is driven by improvements in the experimental
precision.
\begin{table}
\begin{tabular}{|c|c|c|} \hline
Theory includes: & Effects on observable: & Experimental precision: \\ \hline
final-state QED &  shift in $M_W$:  &  Tevatron RUN I: \\ 
(approximation)~\cite{Berends:1984xv} &  
-65$\pm$ 20 (-168$\pm$ 20 ) MeV in the $e(\mu)$ case & $\delta M_W^{exp.} = 59$ MeV  \\ 
&  & $\delta \Gamma_W^{exp.} = 87$ MeV  \\ \hline
full EW ${\cal O}(\alpha)$ 
contribution to resonant & shift in $M_W$: &  Tevatron RUN II:\\
$W$ production (pole approx.)~\cite{Wackeroth:1996hz,Baur:1998kt} & $\approx 10$ MeV & $\delta M_W^{exp.} = 27$ MeV \\ \hline
full EW ${\cal O}(\alpha)$ 
 corrections & affects distributions at high $Q^2$ and & 
Tevatron RUN II: \\
& direct $\Gamma_W$ measurement, shift in $\Gamma_W$: $\approx$ 7 MeV~\cite{Baur:2004ig} &  $\delta \Gamma_W^{exp.} = 25-30$ MeV\\ \hline
real two-photon radiation~\cite{Baur:1999hm} & significantly changes shape of $M_T$ & \\ \hline
multiple final-state photon radiation & shift in $M_W$: & LHC: \\ 
& $2 (10)$ MeV in the $e(\mu)$ case~\cite{CarloniCalame:2003ux} & $\delta M_W^{exp.}$=15 MeV\\ \hline
\end{tabular} 
\caption{Impact of EW radiative corrections on $W$ boson observables,
in particular $M_W$ and $\Gamma_W$ extracted from the 
$M_T(l\nu)$ distribution, confronted with present and anticipated experimental
accuracies~\cite{Baur:2002gp,Abe:1995np,Abachi:1996ey,tevewwg,Abazov:2003sv}.}
\label{tab:b}
\end{table}

\section{Work in Progress}
The EW working group of the TeV4LHC workshop~\footnote{See {\tt http://conferences.fnal.gov/tev4lhc} for more information.}. is
presently addressing the following questions: What is the residual
theoretical uncertainty of the best, presently available predictions
for $W$ and $Z$ boson production at hadron colliders ?  Do we need more
theoretical improvements to be able to fully exploit the EW physics
potential of the Tevatron and the LHC ?  As a first step, the EW working group 
will perform a tuned numerical comparison of available codes that provide
precise predictions for $W$ and $Z$ observables (see
Table~\ref{tab:a}) in the spirit of the LEPI/II CERN yellow books.
First results of a tuned comparison of $W$ and $Z$ production cross
sections and kinematic distributions can be found in
Ref.~\cite{Buttar:2006zd}.  First studies of effects of combined EW
and QCD corrections~\cite{Cao:2004yy}, higher-order EW Sudakov-like
logarithms and multiple final-state photon radiation suggest that for
the anticipated precision at the LHC these effects need to be included
in the data analysis.  Moreover, the model for non-perturbative QCD
contributions~\cite{Konychev:2005iy}, small $x$
effects~\cite{Berge:2004nt} and the impact of heavy-quark
masses~\cite{Berge:2005rv} need to be well understood for a
detailed description of the $p_T(W)$ distribution.  Several groups are
presently working on the combination of EW and QCD radiative
corrections in one MC program, the interface of higher-order EW
calculations, i.e. multiple photon radiation from final-state leptons
and EW Sudakov logarithms, with fixed ${\cal O}(\alpha)$ calculations,
and the calculation of mixed QED/QCD two-loop corrections of ${\cal
O}(\alpha \alpha_s)$, which are not yet available.  The ultimate goal
is to provide one unified MC program that includes all relevant QED, EW
and QCD radiative corrections to $W$ and $Z$ boson production that
matches the anticipated experimental capabilities of the Tevatron and
LHC for EW precision physics.







\section{Conclusions}
Electroweak gauge boson physics offers plentiful and unique
opportunities to test the SM and search for signals of physics beyond
the SM.  Impressive progress has been made in providing precise
predictions at NLO, NNLO QCD and NLO EW and higher (in leading
logarithmic approximation), and a number of MC programs have been made
available to study their effects on observables.  In the context of
the TeV4LHC workshop, these tools are used to determine if they are
sufficient in view of the anticipated experimental capabilities for EW
precision physics at the Tevatron and the LHC. There is ongoing work
on further improving the predictions for $W$ and $Z$ boson observables
and on providing one MC program, including all relevant QED, EW and QCD
corrections, which will meet the high standards of $W$ and $Z$ boson
measurements at the Tevatron and the LHC.

\vspace*{-0.2cm}
\begin{theacknowledgments}
I would like to thank the organizers of the Hadron Collider Physics
Symposium 2006 for the invitation and a very interesting meeting.  My
work is supported in part by the National Science Foundation
under grants NSF-PHY-0244875 and NSF-PHY-0547564, and by the
U.S. Department of Energy under grant DE-FG03-93ER40757.
\end{theacknowledgments}



\bibliographystyle{aipproc}   



\end{document}